\documentclass[conference]{IEEEtran}
\IEEEoverridecommandlockouts
\usepackage{cite}
\usepackage{amsmath,amssymb,amsfonts}
\usepackage{algorithmic}
\usepackage{graphicx}
\usepackage{textcomp}
\usepackage{xcolor}
\usepackage{bm}
\usepackage{subfigure}
\usepackage{url}
\def\BibTeX{{\rm B\kern-.05em{\sc i\kern-.025em b}\kern-.08em
    T\kern-.1667em\lower.7ex\hbox{E}\kern-.125emX}}

\begin{document}

\title{CTT-Net: A Multi-view Cross-token Transformer for Cataract Postoperative Visual Acuity Prediction\\

\thanks{\IEEEauthorrefmark{2}These authors contributed equally to this work. \IEEEauthorrefmark{1}To whom correspondence should be addressed. The copyright notice is: 978-1-6654-6819-0/22/\$31.00 ©2022 IEEE}
}

\author{
\IEEEauthorblockN{
Jinhong Wang\raisebox{0pt}[0pt][0pt]{\textsuperscript{\footnotesize\ensuremath{1}}}\IEEEauthorrefmark{2},
Jingwen Wang\raisebox{0pt}[0pt][0pt]{\textsuperscript{\footnotesize\ensuremath{2}}}\IEEEauthorrefmark{2},
Tingting Chen\raisebox{0pt}[0pt][0pt]{\textsuperscript{\footnotesize\ensuremath{1}}},
Wenhao Zheng\raisebox{0pt}[0pt][0pt]{\textsuperscript{\footnotesize\ensuremath{1}}},
Zhe Xu\raisebox{0pt}[0pt][0pt]{\textsuperscript{\footnotesize\ensuremath{2}}}, \\
Xingdi Wu\raisebox{0pt}[0pt][0pt]{\textsuperscript{\footnotesize\ensuremath{2}}},
Wen Xu\raisebox{0pt}[0pt][0pt]{\textsuperscript{\footnotesize\ensuremath{2}}}\IEEEauthorrefmark{1},
Haochao Ying\raisebox{0pt}[0pt][0pt]{\textsuperscript{\footnotesize\ensuremath{3,4}}}\IEEEauthorrefmark{1},
Danny Chen\raisebox{0pt}[0pt][0pt]{\textsuperscript{\footnotesize\ensuremath{5}}}, and
Jian Wu\raisebox{0pt}[0pt][0pt]{\textsuperscript{\footnotesize\ensuremath{6}}}}
\IEEEauthorblockA{\raisebox{0pt}[0pt][0pt]{\textsuperscript{\footnotesize\ensuremath{1}}}College of Computer Science and Technology, Zhejiang University, Hangzhou, China}
\IEEEauthorblockA{\raisebox{0pt}[0pt][0pt]{\textsuperscript{\footnotesize\ensuremath{2}}}Eye Center of the Second Affiliated Hospital, School of Medicine, Zhejiang University, Hangzhou, China}
\IEEEauthorblockA{\raisebox{0pt}[0pt][0pt]{\textsuperscript{\footnotesize\ensuremath{3}}}School of Public Health, Zhejiang University, Hangzhou, China}
\IEEEauthorblockA{\raisebox{0pt}[0pt][0pt]{\textsuperscript{\footnotesize\ensuremath{4}}}Key Laboratory of Intelligent Preventive Medicine of Zhejiang Province, Hangzhou, China}
\IEEEauthorblockA{\raisebox{0pt}[0pt][0pt]{\textsuperscript{\footnotesize\ensuremath{5}}}University of Notre Dame, USA}
\IEEEauthorblockA{\raisebox{0pt}[0pt][0pt]{\textsuperscript{\footnotesize\ensuremath{6}}}Second Affiliated Hospital School of Medicine, School of Public Health, \\ and Institute of Wenzhou, Zhejiang University, Hangzhou, China}

\IEEEauthorblockA{\{wangjinhong, 3150104381, trista\_chen0603, zhengwenhao, x\_z\_dahaizhe, \\12118542, xuwen2003, haochaoying, wujian2000\}@zju.edu.cn, \{dchen\}@nd.edu}
}

\maketitle

\begin{abstract}
Surgery is the only viable treatment for cataract patients with visual acuity (VA) impairment. Clinically, to assess the necessity of cataract surgery, accurately predicting postoperative VA before surgery by analyzing multi-view optical coherence tomography (OCT) images is crucially needed. Unfortunately, due to complicated fundus conditions, determining postoperative VA remains difficult for medical experts. Deep learning methods for this problem were developed in recent years. Although effective, these methods still face several issues, such as not efficiently exploring potential relations between multi-view OCT images, neglecting the key role of clinical prior knowledge (e.g., preoperative VA value), and using only regression-based metrics which are lacking reference. In this paper, we propose a novel Cross-token Transformer Network (CTT-Net) for postoperative VA prediction by analyzing both the multi-view OCT images and preoperative VA. To effectively fuse multi-view features of OCT images, we develop cross-token attention that could restrict redundant/unnecessary attention flow. Further, we utilize the preoperative VA value to provide more information for postoperative VA prediction and facilitate fusion between views. Moreover, we design an auxiliary classification loss to improve model performance and assess VA recovery more sufficiently, avoiding the limitation by only using the regression metrics. To evaluate CTT-Net, we build a multi-view OCT image dataset collected from our collaborative hospital. A set of extensive experiments validate the effectiveness of our model compared to existing methods in various metrics. Code is available at: \url{https://github.com/wjh892521292/Cataract_OCT}.
\end{abstract}


\vspace{-2pt}
\noindent
\section{Introduction}
Cataract is a very common eye 
disease, often occurring in the elderly~\cite{hoffer1980biometry} and possibly leading to vision diminution and even blindness~\cite{zhu2018dna,holden2016global}.
The only effective treatment for this public health problem is surgery~\cite{thompson2015cataracts}. Clinically, evaluating postoperative visual acuity (VA) can provide a direct reference to estimate the necessity and effectiveness of cataract surgery, and postoperative VA is assessed mainly by the transparency of refractive media and retinal fundus status in multi-view (horizontal and vertical views) optical coherence tomography (OCT) images~\cite{huang2018macular,li2018paravascular}. However, due to complicated fundus conditions~\cite{li2018paravascular,chang2013myopia,todorich2013macular,gohil2015myopic,lichtwitz2016prevalence} (e.g., chorioretinal atrophy, foveoschisis), this morphology task is laborious and error-prone, even for experienced doctors. Thus, it is desirable to develop computer-aided methods to predict postoperative VA before surgery.

\begin{figure*}[ht]
\centering
{
  \includegraphics[width=0.20\textwidth]{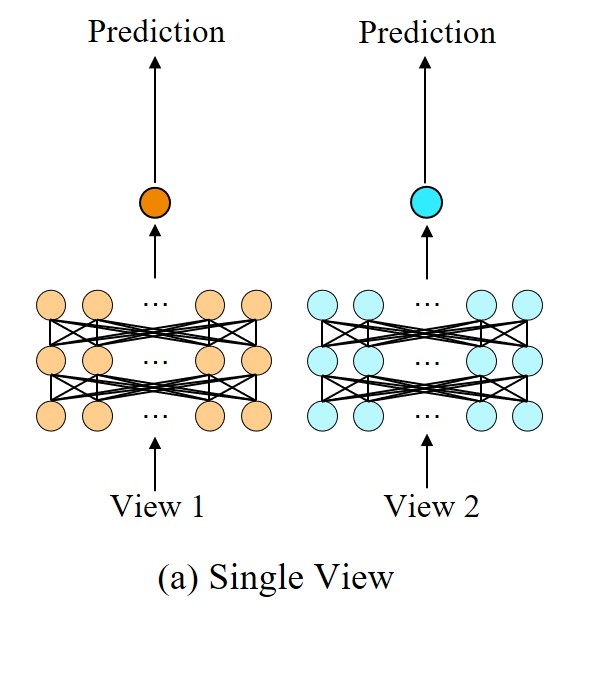}%
}\hfil
{%
  \includegraphics[width=0.20\textwidth]{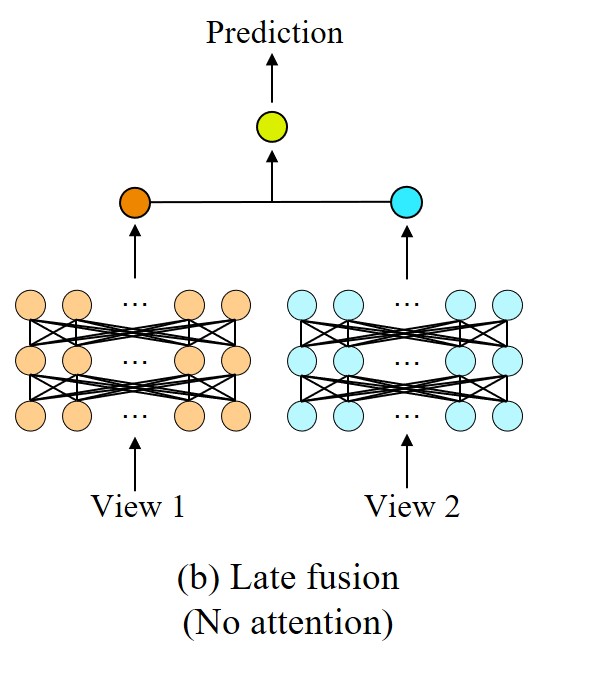}%
}\hfil
{%
  \includegraphics[width=0.20\textwidth]{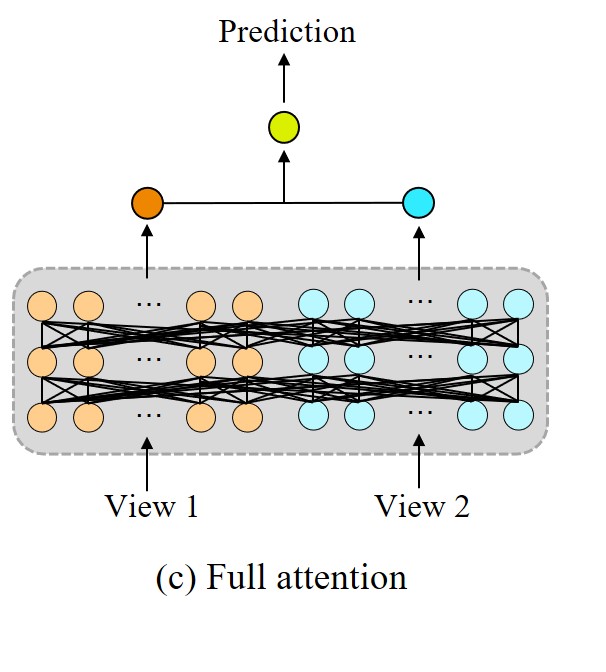}%
}\hfil
{%
  \includegraphics[width=0.20\textwidth]{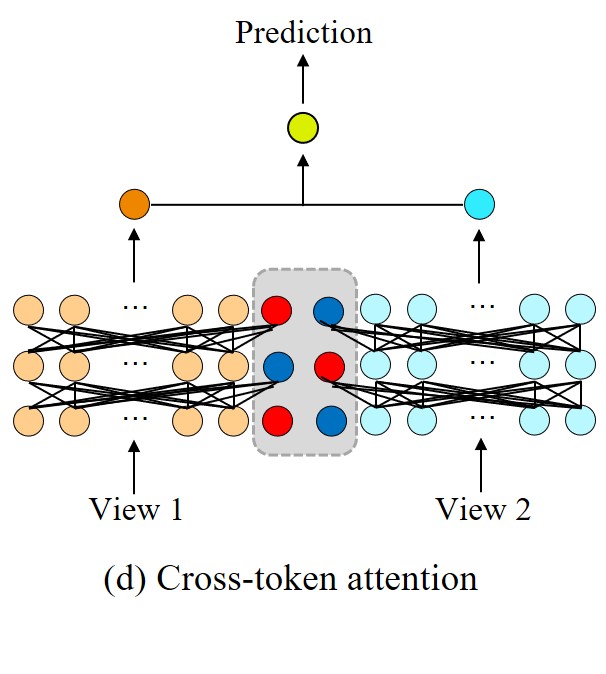}%
}
\vspace{-6pt}
\noindent
\caption{Four types of multi-view fusion methods. (a) Simply making separate predictions. (b) Extracting features of each view separately, and averaging them without any attention flow between views. (c) Combining features using full attention between views. (d) Our sectional fusion: combining features using cross-token attention between views. Grey boxes mark tokens that receive attention flow from tokens of each view.}
\label{fig1}
\vspace{-6pt}
\end{figure*}
Exploiting powerful representation capabilities of deep learning (DL) on medical image data, known DL-based methods have been developed for this task. For instance, by adapting ResNet~\cite{he2016deep} and Inception~\cite{szegedy2016rethinking} to extract features of OCT images, Ling et al.~\cite{wei2021optical} analyze single view images to predict postoperative VA as Fig.~\ref{fig1}(a). But using only single view images to predict does not consider feature correlations between multi-view images, and can lose some potential and useful information. In particular, some previous studies combine features of multi-view OCT images by late fusion~\cite{lichtwitz2016prevalence,yoo2021deep}, that is, a model first extracts features of the two views independently, and then concatenates their features and outputs the final prediction by a fully connected (FC) layer. Since there is no attention flow between views for fusion before the final FC layer, these late-fusion methods (Fig.~\ref{fig1}(b)) only focus on semantic information, but neglect to capture dependencies between low-level features of views with rich spatial information. Unlike these non-attention methods, a transformer structure~\cite{otsuki2022integrating} fuses low-level features by allowing full attention flow (Fig.~\ref{fig1}(c)) between features of different views. Although effective, the full-attention transformer faces a crucial issue in high density and redundancy using pairwise attention flow. Thus, for attention flow constraint and effective fusion, multi-view fusion methods need to be further explored.

In addition, known postoperative VA prediction methods also ignore two important aspects considered in clinics.
First, known methods do not consider useful information from other clinical data. For instance, preoperative VA, as a key role of clinical prior knowledge, could relate directly to image features by reflecting intuitive assessments of fundus states and probabilities of other eye diseases~\cite{faulkner1983laser,uy2005comparison,obata2021prediction}. Thus, to enhance image feature fusion and provide more information and reference for postoperative VA prediction, such profitable information should be exploited to train DL models. Second, known models are all trained with regression metrics such as MAE and RMSE. These metrics are not meaningful enough since predicting specific postoperative VA is often biased due to all sorts of complications, and considering that doctors can assess the necessity of surgery based on whether the postoperative VA recovery is better than a certain standard~\cite{gohil2015myopic}, these regression metrics are not sufficiently determining the possibility of VA recovery. To avoid these limitations and better measure the necessity of the surgery, it is essential to explore other more comprehensive and referable metrics.

To tackle the aforementioned issues, we propose a novel multi-view fusion model, Cross-Token Transformer Network (CTT-Net), for cataract postoperative VA prediction. We develop cross-token attention(Fig.~\ref{fig1}(d)) to fuse the horizontal and vertical views features of OCT images efficiently and effectively. The cross-token attention forces the model to collate features of each view into cross-tokens and exchange them between views to restrict redundant/unnecessary attention flow. Moreover, we add preoperative VA as supplementary information to provide more guidance for postoperative VA prediction and facilitate multi-view fusion in transformer. Further, we design an auxiliary classification loss for an additional classification task to distinguish whether the VA increase after surgery is bigger than our preset threshold ($0.2$ in clinical) so that our model can avoid the limitations of regression metrics used only and better help doctors evaluate the necessity of surgery. 
We build a multi-view OCT image dataset collected from our collaborative hospital which contains 1393 samples. Extensive experiments validate the superiority of our model compared to existing methods in various metrics and the effectiveness of each component of our method.

\section{THE PROPOSED CTT-NET}
\vspace{-10pt}
\noindent
\subsection{Overview}
\begin{figure*}[ht]
\centering
\subfigure[CTT-Net]{%
    \label{fig3:subim1}
  \includegraphics[width=0.5\textwidth]{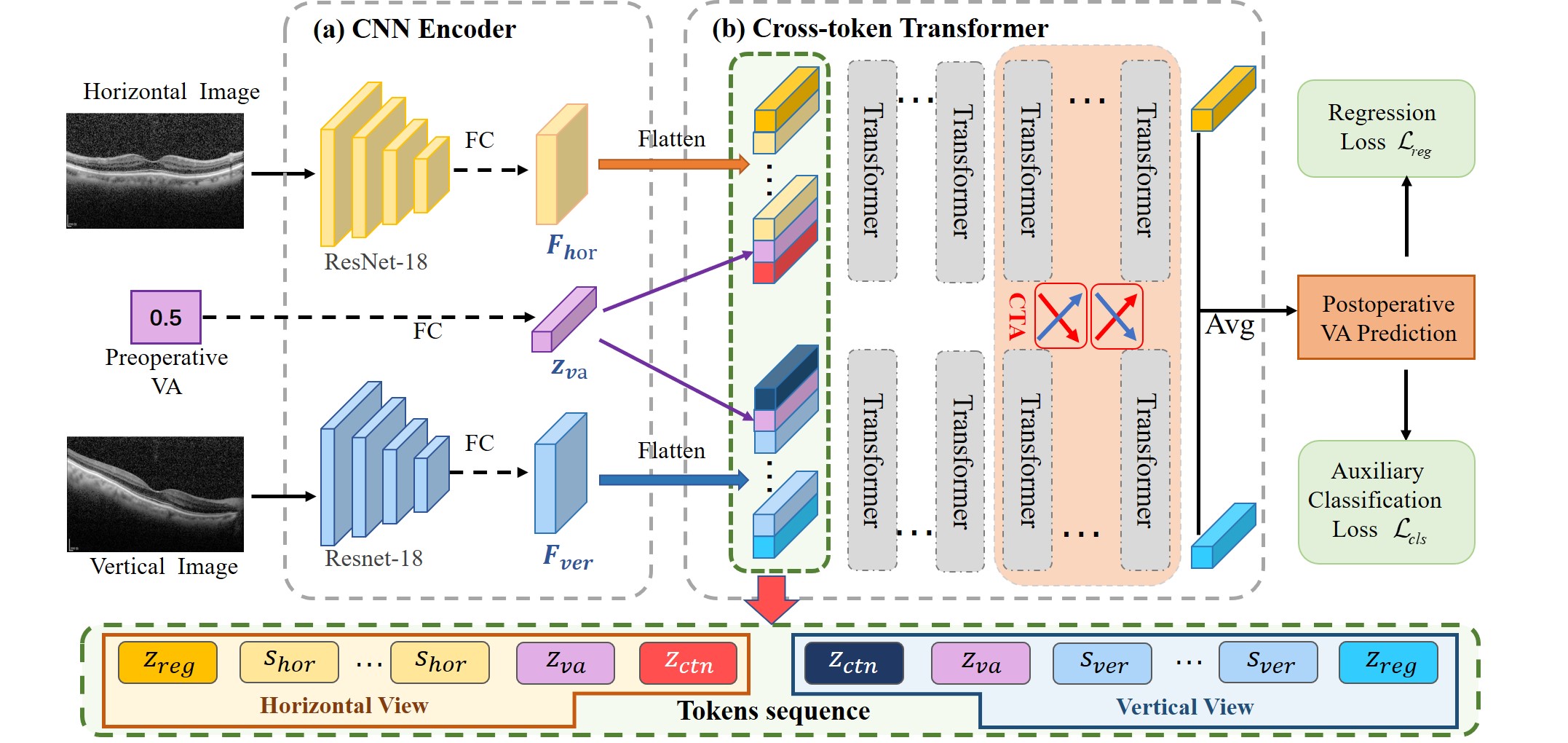}%
 
}\hfil
\subfigure[Cross-token Attention]{%
    \label{fig3:subim2}
  \includegraphics[width=0.45\textwidth]{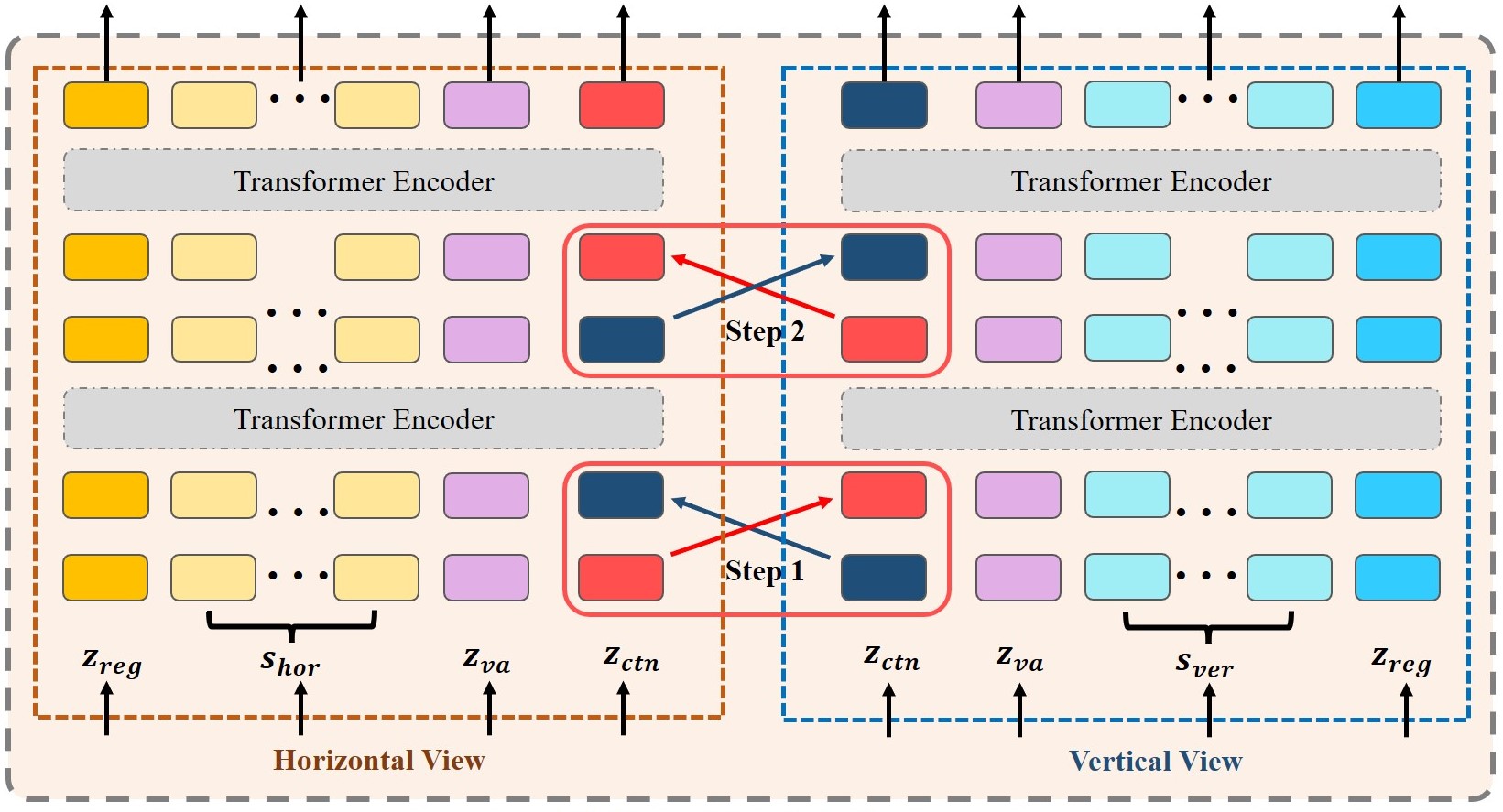}%
 
}
\vspace{-6pt}
\noindent
\caption{The overview of CTT-Net (a) and Cross-token Attention (b).}
\label{fig2}
\vspace{-6pt}
\end{figure*}

Fig.~\ref{fig3:subim1} shows an overview of our CTT-Net. Given horizontal and vertical OCT images, and a preoperative VA value, our CTT-Net predicts the postoperative VA value by two main parts. One part is a CNN encoder for extracting and encoding images features into a token-based sequence. The other part is the Cross-token Transformer which exerts the important role of preoperative VA and explores potential relations between features of different views. Our CTT-Net is trained with a common regression loss and a novel auxiliary classification loss to measure the postoperative VA recovery in a clinical and more comprehensive way. The details are given below.

\vspace{-6pt}
\noindent
\subsection{CNN Encoder}
As shown in Fig.~\ref{fig2}(a), the CNN Encoder has a feature extractor for the OCT image of each view. Encouraged by Ling's work~\cite{wei2021optical}, we use ResNet-18 as our feature extractor, and the input of each ResNet-18 is a grayscale image of size $H \times W$. Encoded by four layers of ResNet-18 and a fully connected (FC) layer, the obtained feature maps of horizontal and vertical views are denoted by $F_{hor}, F_{ver} \in \mathbb{R} ^ {P_{1} \times P_{2} \times D}$, respectively, where $P_{1}=\frac{H}{32}$, $P_{2}= \frac{W}{32}$, and $D$ is a dimension of the feature maps that is equal to an input dimension of the Cross-token Transformer. Since the Transformer takes a sequence of 1D vectors as input, we arrange the feature maps of each view into a token sequence, $s_{hor}/ s_{ver} \in \mathbb{R} ^ {P \times D}$, where $P=P_{1} \times P_{2}$. Meanwhile, a preoperative VA is converted into a token $z_{va} \in \mathbb{R} ^ {1 \times D}$ by an FC layer, as a feature supplement that concatenated with imaging tokens. In the Transformer, information will transmit between tokens so that the preoperative VA token can enhance the features both in horizontal and vertical views for better multi-view fusion, and provide more information for postoperative VA prediction.

\vspace{-6pt}
\noindent
\subsection{Cross-token Transformer}
After tokenization, the Cross-token Transformer utilizes transformer layers with cross-token attention to explore dependency and relations between tokens for multi-view information integration and postoperative VA prediction(see Fig.~\ref{fig3:subim2}). Our Cross-token Transformer contains $L$ transformer encoder layers, and we adopt cross-token attention only in late layers to perform the multi-view fusion. In this way, our model can learn and summarize information from imaging and VA tokens of each view at the early layers, and then effectively fuse multi-view features using cross-token attention.

\paragraph {Transformer Encoder} For each view (e.g., the horizontal view), the input token sequence of the transformer encoder can be described as follows:
\vspace{-6pt}
\noindent
\begin{equation}
     {\rm z_{ipt_{hor}}}= [z_{reg_{hor}}, z_{va}, s^1_{hor}, s^2_{hor},\ldots, s^P_{hor}, z_{ctn_{hor}}],
\end{equation}
where $z_{reg} \in \mathbb{R}^{ 1 \times D }$ is a special token prepended to this sequence so that its representation at the final layer can be passed to a regressor for a regression task, and ${\rm z_{ctn}} \in \mathbb{R}^{ 1 \times D }$ is a cross-token for Cross-token attention. Formally, the input token sequence ${\rm z_{ipt_{ver}}}$ for vertical view is similar.  Each transformer encoder layer contains Multi-Headed Self-Attention (MSA), Layer Normalisation (LN), and Feed-Forward Network (FFN) blocks using residual connections. We denote a transformer encoder layer, ${\rm z}^{l+1} =$ Transformer$({\rm z}^l)$ as:
\vspace{-3pt}
\noindent
\begin{align}
    {\rm y}^l = {\rm MSA}({\rm LN}({\rm z}^{l})) + {\rm z}^l,\\
     {\rm z}^{l+1} = {\rm FFN}({\rm LN}({\rm y}^{l})) + {\rm y}^l,
     \vspace{-6pt}
\end{align}
\noindent
here ${\rm z}^l$ denotes the output of the $l^{th}$ transformer encoder layer. 


 
\vspace{2pt}
\noindent
\paragraph {Cross-token Attention} The cross-tokens learned global information of each view in early transformer layers, and to fuse multi-view global information effectively, we propose cross-token attention applied in the remaining layers.  Cross-token attention only exchanges those cross-tokens between different views which avoids pairwise attention to limited computational cost. For convenience, we define ${\rm z_{hor}} = [z_{reg_{hor}}, z_{va}, s^1_{hor}, s^2_{hor}, \ldots, s^P_{hor}]$ to represent the unchanged tokens for horizontal view, and the ${\rm z_{ver}}$ for vertical view is similar.
Then the $l^{th}$ layer of the Transformer processes as:
\vspace{-3pt}
\noindent
\begin{align}
    [{\rm z}^{l+1}_{\rm hor} || {\rm z}^{l+1}_{\rm ctn_{hor}}] &= {\rm Transformer}([{\rm z}^{l}_{\rm hor} || {\rm z}^{l}_{\rm ctn_{hor}}]; \theta_{hor}),\\
    [{\rm z}^{l+1}_{\rm ver} || {\rm z}^{l+1}_{\rm ctn_{ver}}] &= {\rm Transformer}([{\rm z}^{l}_{\rm ver} || {\rm z}^{l}_{\rm ctn_{ver}}]; \theta_{ver}),
\end{align}
where $\theta_{hor}$ and $\theta_{ver}$ are model parameters ($\theta_{hor}$ and $\theta_{ver}$ are not equal).  The model can gradually fuse multi-view features by exchanging cross-tokens back and forth (as shown in Fig.~\ref{fig2}(b)). Specifically, at each layer, Cross-token attention performs two steps. The first step changes the cross-tokens of different views and forms new sequences to Then the $l^{th}$ layer of the Transformer processes as follows: 
\vspace{-3pt}
\noindent
\begin{align}
[{\rm \hat z}^{l}_{\rm hor} || {\rm \hat z}^{l}_{\rm ctn_{ver}}] &= {\rm Transformer}([{\rm z}^{l}_{\rm hor} || {\rm z}^{l}_{\rm ctn_{ver}}]; \theta_{hor}),\\
[{\rm \hat z}^{l}_{\rm ver} || {\rm \hat z}^{l}_{\rm ctn_{hor}}] &= {\rm Transformer}([{\rm z}^{l}_{\rm ver} || {\rm z}^{l}_{\rm ctn_{hor}}]; \theta_{ver}),
\end{align}
where  $[{\rm \hat z}^{l}_{\rm hor} || {\rm \hat z}^{l}_{\rm ctn_{ ver}}]$ and $[{\rm \hat z}^{l}_{\rm ver} || {\rm \hat z}^{l}_{\rm ctn_{hor}}]$ are temporary token sequences for each view.
The second step changes the cross-tokens back, in order to integrate information and restore token sequences, as follows:
\vspace{-6pt}
\noindent
\begin{align}
    [{\rm z}^{l+1}_{\rm hor} || {\rm z}^{l+1}_{\rm ctn_{hor}}] &= {\rm Transformer}([{\rm \hat z}^{l}_{\rm hor} || {\rm \hat z}^{l}_{\rm ctn_{hor}}]; \theta_{hor}),\\
    [{\rm z}^{l+1}_{\rm ver} || {\rm z}^{l+1}_{\rm ctn_{ver}}] &= {\rm Transformer}([{\rm \hat z}^{l}_{\rm ver} || {\rm \hat z}^{l}_{\rm ctn_{ver}}]; \theta_{ver}).
\end{align}
The above strategies are for fusion within a layer, and features of each view can only exchange
information via cross-tokens ${\rm z_{ctn_{hor}}}$ and ${\rm z_{ctn_{ver}}}$. Since all cross-modal attention must pass through cross-tokens, it forces the model to condense information into cross-tokens from each view and fuse information with less attention flow. Then we repeatedly exchange the cross-tokens to fuse multi-view features deeply. And unlike full-attention methods, another advantage of our model is that most tokens are unchanged which can retain the specific features of each view to the greatest extent for maintaining the integrity of the features. 
\vspace{-6pt}
\noindent
\subsection{Loss Functions}
\label{sec-loss}    
\vspace{-6pt}
\noindent

\begin{table*}[ht]
    \centering
    \vspace{-6pt}
    \caption{Model Performance. HOR/VER = Horizontal/Vertical, PVA = Preoperative VA, and ACL = Auxiliary Classification Loss. }
    \vspace{-6pt}

    \label{tab1}
    \scalebox{0.92}{
    \begin{tabular}{|l|l|l|l|l|l|}
        \hline
        Method &  Views & MAE ($\downarrow$) & RMSE ($\downarrow$) & ACC ($\uparrow$) & F1-s ($\uparrow$) \\
        \hline
        Single view~\cite{wei2021optical}& HOR & $0.168\pm  0.014$ & $0.219\pm0.014$ & $0.805\pm0.034$ & $0.866\pm0.022$ \\ 
        
        Single view~\cite{wei2021optical}& VER & $0.167\pm0.013$ & $0.212\pm0.012$ & $0.807\pm0.034$ & $0.870\pm0.021$\\
        
        No attention~\cite{lichtwitz2016prevalence}& HOR+VER & $0.162\pm0.012$ & $0.207\pm0.016$ & $0.821\pm0.033$ & $0.878\pm0.025$\\
      
        Full attention~\cite{otsuki2022integrating}& HOR+VER & $0.157\pm0.014$ & $0.197\pm0.014$ & $0.832\pm0.024$ & $0.883\pm0.021$\\
        \hline
        Cross-token attention (CTA) & HOR+VER &  $0.153\pm0.013$ & $0.195\pm0.016$ & $0.845\pm0.022$ & $0.892\pm0.020$ \\
      
        CTA+PVA &  HOR+VER & $0.145\pm0.012$ & $0.189\pm0.016$ & $0.867\pm0.019$ & $0.909\pm0.017$ \\
      
        {\textbf{CTA+PVA+ACL (CTT-Net)}} & \textbf{HOR+VER} & \bm{$0.144\pm0.012$}& \bm{$0.189\pm0.012$} & \bm{$0.874\pm0.017$} & \bm{$0.917\pm0.016$} \\
        \hline
    \end{tabular}}
    \vspace{-6pt}

\end{table*}

Regression metrics are commonly used in previous methods for postoperative VA evaluation. However, these metrics are not enough to provide a reliable reference since predicting specific postoperative VA is often biased and the probability of good VA recovery after surgery is not reflected. Therefore, to better measure the necessity of the surgery, we add a classification metric to predict whether the VA can recover to a 
healthy level. And we design an auxiliary classification loss to enhance the model performance, especially in classification metrics. Hence, our total loss function contains a regression loss and an auxiliary classification loss.

\subsubsection {Regression Loss} The output of CTT-Net is a postoperative VA value. We choose the mean squared error (MSE) as the regression loss, defined as follows:
\vspace{-6pt}
\noindent
\begin{equation}
    L_{reg} = MSE(\tilde{y_i},y_i)=\frac{1}{N}\sum_{i=1}^{N}(\tilde{y_i}-y_i)^2,
\end{equation}
where $\tilde{y_i}$ is the predicted postoperative VA, ${y_i}$ is the true postoperative VA, and $N$ is total number of samples.

\noindent
\subsubsection {Auxiliary Classification Loss} Aiming toward a better measure for postoperative recovery, we add a metric that represents the accuracy of label pairs ($Y$, $\tilde{Y}$) being equal. $Y$ ($\tilde{Y}$) is 1 if the VA difference between the true (predicted) postoperative VA  and preoperative VA  is $>0.2$; otherwise, it is 0. To enhance the classification metric, we propose an auxiliary classification loss defined as follows:
\vspace{-6pt}
\noindent
\begin{equation}
    L_{cls} = \frac{1}{N}\sum_{i=1}^{N}[Y_{i}Relu(-P(\tilde{y_i})) + ( 1 - Y_{i} )Relu(P(\tilde{y_i}))]
\end{equation}
where $P(\tilde{y_i})$ denotes the penalty weight that is equal to $\tilde{y_i}-x_i - 0.2$, and $x_i$ is the preoperative VA. With the assistance of this loss function, we can effectively improve the accuracy metric by imposing penalties on samples whose predictions are inconsistent with the true labels.

Then, with the regularization
parameter $\lambda$, the total loss is:
\vspace{-6pt}
\noindent
\begin{equation}
    L_{tot} = L_{reg} + \lambda L_{cls}.
\end{equation}

\vspace{-6pt}
\noindent
\section{Experiments}
\vspace{-2pt}
\noindent
\subsubsection{Dataset}
We evaluate the performance of our CTT-Net approach on a dataset of 1393 patients, provided by a local hospital. The dataset contains one horizontal OCT image, one vertical OCT image, and a preoperative VA value for every patient during each visit. The corresponding postoperative VA value is collected one month after surgery. 
\vspace{1pt}
\noindent
\subsubsection {Evaluation Metrics} For all our experiments, we conduct a 5-fold cross-validation, and report the average results and standard deviation. We evaluate the regression performance in mean average error (MAE) and root mean squard error (RMSE). Further, based on the VA labels $Y$ ($\tilde{Y}$) discussed in Sect.~\ref{sec-loss}, we use accuracy (ACC) and F1-score (F1-s) to evaluate the classification performance.

\vspace{1pt}
\noindent
\subsubsection {Implementation} We adopt PyTorch for implementation, and all the models are trained using SGD for optimization with a momentum = 0.9. We train the network for 3000 iterations with a batch size of 16. We use a step-wise learning rate scheduler with an initial learning rate of $1e-3$ and decay by 0.1 every 1000 iterations. For data augmentation, we apply random rotation, random flip, random mirror, random brightness, and random gray-scale. After random augmentation, the images are scaled to size 256 $\times$ 256. We choose ResNet-18 as our CNN encoder and use ImageNet-pretrained weights. In the Cross-token Transformers, we use transformers with 3 attention heads. The dimension $D$ of tokens is set to 128 and the number of layers $L$ of the transformer encoder is set to 12. The first 6 layers do not apply cross-token attention while the last 6 layers do. For the loss function, the regularization parameter $\lambda$ is set to 2. The optimal values of these hyper-parameters are determined by our experimental testing.

\vspace{1pt}
\noindent
\subsubsection {Comparing with Known Methods} We implement and adapt several multi-view fusion methods and one state-of-the-art method in postoperative VA prediction for comparison. (1) Single view~\cite{wei2021optical}: A state-of-the-art method. It uses several CNN backbones such as ResNet-18, ResNet-50, and Inception-V3, to directly predict postoperative VA. (2) No attention: It uses ResNet-18 as CNN encoder and applies the late fusion method by concatenating multi-view features at the last FC layer and average output as the postoperative VA prediction. (3) Full attention: It uses Vit~\cite{dosovitskiy2020image} to fuse multi-view features obtained by ResNet-18 to predict the postoperative VA where the Vit includes 12 layers of a Transformer Encoder. As shown in the first block of Table~\ref{tab1}, no matter which method is used for fusion, the performance of the model using multiple views is generally superior to the model using only one single view. However, different fusion methods lead to different results. It is obvious that the performance improvement of no attention fusion is very limited. We hypothesize that this is because this method fuses features only at the highest level, which includes only rich semantic information while neglecting global feature information at low levels that is of equipollent importance. By applying transformer, the full attention flow of early features can provide spatial information of different views, which is conducive to the overall feature fusion, positioning, and alignment of OCT macular regions, so as to improve the effectiveness of the model. But, full pairwise attention at all the layers of the transformer is in some way excessive, leading to possible loss of its own features in each view. The comparison of experimental results shows that this problem is handled by our cross-token attention verifying that the restriction of redundant attention flow is effective. On the other hand, it also verifies the feasibility of using a single token to condense information of a view and act as a carrier of information interaction and fusion.


\vspace{0pt}
\noindent
\subsubsection {Ablation Study} We conduct an ablation study to examine the effects of our proposed ideas, shown in the second block of Table~\ref{tab1}. The baseline is the Single view model. After adding cross-token attention, all the evaluation metrics improve considerably, implying that reducing redundant attention flow between views using cross-tokens is helpful.  Comparing the performances of CTT-Net with or without preoperative VA, it is obvious that preoperative VA is useful clinical prior knowledge. Also, applying auxiliary classification loss enhances the model performance, especially in the classification metrics. Our model achieves 87.4\%  accuracy and 91.7 \% F1-score in predicting whether patients would recover. These results show that all the modules in our method are effective for postoperative VA prediction and our model can provide more reliable assistant diagnoses.

\begin{figure}[t]
\centering
\vspace{-6pt}
\subfigure[\label{fig4:subim1}]{%
  \includegraphics[width=0.20\textwidth]{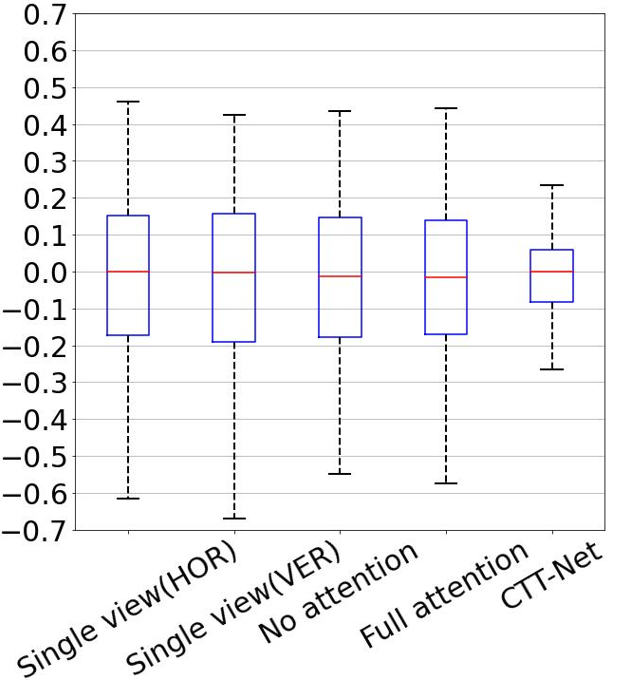}%
}\hfil
\subfigure[\label{fig4:subim2}]{%
  \includegraphics[width=0.20\textwidth]{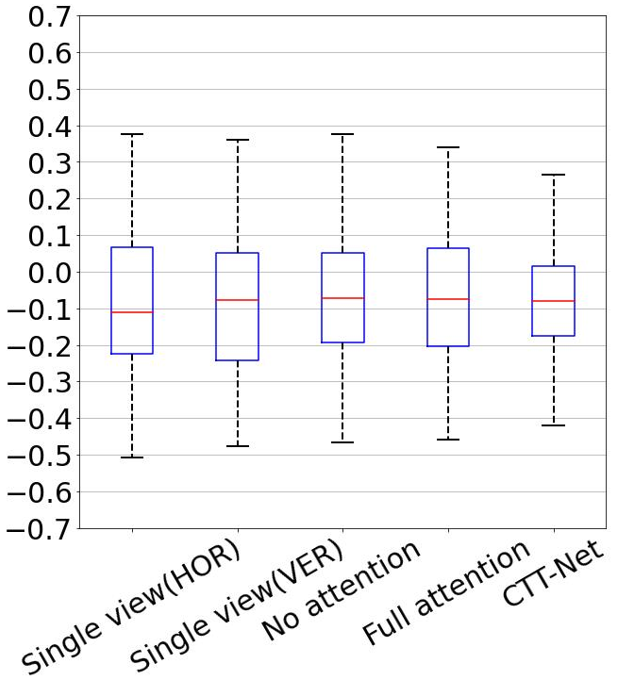}%
}
\vspace{-6pt}
\noindent
\caption{Boxplot of postoperative VA gap for various methods divided into two groups by true postoperative VA: (a) high VA(0.7-1.5); (b) low VA(0-0.7).}
\vspace{-6pt}
\noindent
\label{fig4}
\end{figure}

\vspace{0pt}
\noindent
\subsubsection {Distribution Evaluation} To compare the more detailed performance of different methods, we visualize the distribution of the VA gap (true postoperative VA - predicted postoperative VA) using boxplots as shown in Fig.~\ref{fig4}. The boxplots of the VA gap show the scope of pairwise difference between true label and model prediction which can validate the distribution predicted by the model. According to clinical experience, a patient is considered healthy if true postoperative VA is higher than 0.7. Based on this standard, we divide all samples into two groups where group A includes 187 samples with high VA and group B includes remained 91 samples with low VA.  For those patients with high postoperative VA, that is the true postoperative VA is better than 0.7, the Fig.~\ref{fig4:subim1} shows that our model CTT-Net performs better since the median, Q1, Q3, minimum and maximum are all closed to zero when compared with other models, which indicates that the prediction of our model is more reliable. However, for patients with poor postoperative VA, Fig.~\ref{fig4:subim2} shows that all models are less effective since all the metrics are further from zero than the high VA group, which we speculate it is attributed to the unbalanced data
distribution. Then we compare all models in the low VA group, our model also achieves better performance which verifies that our proposed method is able to provide more reliable assistant diagnoses for doctors when the data distribution is unbalanced.

\section{Conclusions}
In this paper, we proposed a novel multi-view DL framework for VA prediction after cataract surgery. By tokenization, preoperative VA is input into different views to strengthen the connection and fusion between views. By introducing Cross-token attention, our model well explores inherent correlations between two different OCT image views with higher efficiency. Further, by developing a new auxiliary classification loss, the accuracy of postoperative recovery prediction is directly facilitated. Experiments on a dataset of 1393 pairs of eyes validated the effectiveness of CTT-Net. In future work, we will transfer our methods to 
other types of disease diagnosis to further validate the
effectiveness of CTT-Net.

\section*{Acknowledgment}
\vspace{-2pt}
\noindent
This research was partially supported by National Key R\&D Program of China under grant No. 2019YFC0118802, National Natural Science Foundation of China under grants No. 62176231 and 62106218.

\bibliographystyle{IEEEtran}
\bibliography{refs}

\end{document}